\input harvmac
\input epsf
\noblackbox
\def\tef{t_F}\def\teb{t_B}\def\tee{t_E}
\def\tfp{{t^\prime_F}}\def\tbp{{t^\prime_B}}\def\tep{{t^\prime_E}}

\def\Bp{\cx B^\prime}

\def\us#1{\underline{#1}}
\def\hth/#1#2#3#4#5#6#7{{\tt hep-th/#1#2#3#4#5#6#7}}
\def\nup#1({Nucl.\ Phys.\ $\us {B#1}$\ (}
\def\plt#1({Phys.\ Lett.\ $\us  {B#1}$\ (}
\def\cmp#1({Comm.\ Math.\ Phys.\ $\us  {#1}$\ (}
\def\prp#1({Phys.\ Rep.\ $\us  {#1}$\ (}
\def\prl#1({Phys.\ Rev.\ Lett.\ $\us  {#1}$\ (}
\def\prv#1({Phys.\ Rev.\ $\us  {#1}$\ (}
\def\mpl#1({Mod.\ Phys.\ Let.\ $\us  {A#1}$\ (}
\def\ijmp#1({Int.\ J.\ Mod.\ Phys.\ $\us{A#1}$\ (}
\def\tit#1|{{\it #1},\ }
\def\coeff#1#2{\relax{\textstyle {#1 \over #2}}\displaystyle}
\def\ch{\hat{c}}
\def\subsubsec#1{\ \br \noindent {\it #1}}
\def\IP{{\bf P}}\def\IC{{\bf C}}
\def\br{\hfill\break}
\def\cx#1{ {\cal #1}}

\def\ep{\epsilon}

\def\x{{\bf X}}

\def\h{{1 \over 2}}
\def\IR{ {\bf R}}

\def\IZ{ {\bf Z}}
\def\IF{{\bf F}}
\def\IS{{\bf S}}
\def\del{\Delta}

\def\ca#1{{\cal Q}^{(#1)}}
\def\gc#1{\gamma^{(#1)}}

\def\IR{ {\bf R}}

\def\IZ{ {\bf Z}}
\def\del{\Delta}

\def\ca#1{{\cal Q}^{(#1)}}
\def\gc#1{\gamma^{(#1)}}

\def\IT{{\bf T}}


\def\bfone{\relax{\rm 1\kern-.35em 1}}
\def\inbar{\vrule height1.5ex width.4pt depth0pt}
\def\IC{\relax\,\hbox{$\inbar\kern-.3em{\mss C}$}}
\def\ID{\relax{\rm I\kern-.18em D}}
\def\IH{\relax{\rm I\kern-.18em H}}
\def\II{\relax{\rm I\kern-.17em I}}
\def\IN{\relax{\rm I\kern-.18em N}}
\def\IQ{\relax\,\hbox{$\inbar\kern-.3em{\rm Q}$}}
\def\us#1{\underline{#1}}
\def\IR{\relax{\rm I\kern-.18em R}}
\font\cmss=cmss10 \font\cmsss=cmss10 at 7pt
\def\ZZ{\relax\ifmmode\mathchoice
{\hbox{\cmss Z\kern-.4em Z}}{\hbox{\cmss Z\kern-.4em Z}}
{\lower.9pt\hbox{\cmsss Z\kern-.4em Z}}
{\lower1.2pt\hbox{\cmsss Z\kern-.4em Z}}\else{\cmss Z\kern-.4em
Z}\fi}
\def\a{\alpha} \def\b{\beta}

 \def\cB{{\cal B}}
 
\def\cF{{\cal F}}

 \def\cM{{\cal M}}
 \def\cO{{\cal O}}

\def\nup#1({Nucl.\ Phys.\ $\us {B#1}$\ (}
\def\plt#1({Phys.\ Lett.\ $\us  {B#1}$\ (}
\def\cmp#1({Comm.\ Math.\ Phys.\ $\us  {#1}$\ (}
\def\prp#1({Phys.\ Rep.\ $\us  {#1}$\ (}
\def\prl#1({Phys.\ Rev.\ Lett.\ $\us  {#1}$\ (}
\def\prv#1({Phys.\ Rev.\ $\us  {#1}$\ (}
\def\mpl#1({Mod.\ Phys.\ Let.\ $\us  {A#1}$\ (}
\def\ijmp#1({Int.\ J.\ Mod.\ Phys.\ $\us{A#1}$\ (}
\def\tit#1|{{\it #1},\ }
\def\Coeff#1#2{{#1\over #2}}
\def\Coe#1.#2.{{#1\over #2}}
\def\coeff#1#2{\relax{\textstyle {#1 \over #2}}\displaystyle}
\def\coe#1.#2.{\relax{\textstyle {#1 \over #2}}\displaystyle}

\def\del{\partial}
\def\nex#1{$N\!=\!#1$}

\def\doubref#1#2{\refs{{#1},{#2}}}

\def\br{\hfill\break}
\def\abstract#1{
\vskip .5in\vfil\centerline
{\bf Abstract}\penalty1000
{{\smallskip\ifx\answ\bigans\leftskip 2pc \rightskip 2pc
\else\leftskip 5pc \rightskip 5pc\fi
\noindent\abstractfont \baselineskip=12pt
{#1} \smallskip}}
\penalty-1000}

\def\ca{c}
\def\u{u}
\def\IB{{\bf B}}\def\xf{\x_{\IF_1}}
\def\phit{\tilde{\phi}}
\def\phitd{\tilde{\phi}_D}
\def\ex#1{{}}
\def\nihil#1{{\it #1}}
\def\eprt#1{#1}

\lref\SaitoES{K.\ Saito,
{\it Einfach-Elliptische Singularit\"aten},
Inv.\ Math.\ 23 (1974) 289.}

\lref\HM{J.\ Harvey and G.\ Moore,
\nup 463 (1996) 315; \nihil{On the Algebras of BPS States,}
 \eprt{hep-th/9609017}.}

\lref\cand{P. Candelas, X. De la Ossa, A. Font, S. Katz, and D.
Morrison, \nup429 (1994) 626}

\lref\bcov{M. Bershadsky, S. Ceccotti, H. Ooguri and C. Vafa
(App. by S. Katz), \nup405 (1993) 279,
\cmp165 (1994) 311}

\lref\cvI{C. Vafa, \nup477 (1995) 262}

\lref\gms{O.J. Ganor, D.R. Morrison and N. Seiberg, {\it
Branes, Calabi-Yau Spaces, and Toroidal Compactification of
  the N=1 Six-Dimensional $E_8$ Theory},
hep-th/9610251}

\lref\fer{I. Antoniadis, S. Ferrara and T.R. Taylor, \nup 460 (1996)
489}

\lref\cerI{A. Ceresole, R. D'Auria, S. Ferrara
and A. Van                  Proeyen, \nup 444 (1995) 92}

\lref\seibI{N. Seiberg, \plt 388 (1996) 753}

\lref\dkv{ M. Douglas, S. Katz and C. Vafa, {\it
Small Instantons, Del Pezzo Surfaces and Type I' theory},
hep-th/9609071}

\lref\ms{D. R. Morrison and N. Seiberg, {\it
Extremal Transitions and Five-Dimensional Supersymmetric
Field Theories}, hep-th/9609070}

\lref\chone{S. Mukhi and C. Vafa, \nup 407 (1993) 667}

\lref\ovI{H. Ooguri and C. Vafa, \nup 463 (1996) 55}

\lref\mv{D. R. Morrison and C. Vafa,
\nup 473 (1996) 74, \nup 476 (1996) 437}

\lref\candI{P. Candelas and A. Font, {\it
Duality Between Webs of Heterotic and Type II Vacua},
hep-th/9603170}

\lref\witm{E. Witten, \nup471 (1996) 195}

\lref\witx{P. C. Argyres, M. R. Plesser, N. Seiberg
                  and E. Witten,
\nup461 (1996) 71}

\lref\ste{J. Louis, J. Sonnenschein, S. Theisen and S.
                  Yankielowicz,
\nup 480 (1996) 185}

\lref\eestring{O. J. Ganor and A. Hanany,
\nup 474 (1996) 122}

\lref\ganorI{O. J. Ganor, \nup 479 (1996) 197}

\lref\witnew{E. Witten, {\it Physical Interpretation Of Certain
Strong Coupling Singularities},
hep-th/9609159 }

\lref\gun{M. G\"unaydin, G. Sierra and P.K. Townsend, \nup 242 (1994)
244; \nup 253 (1985) 573}

\lref\hsh{P. Claus et. al., \plt 373 (1996) 81}

\lref\sp{P. Mayr, {\it
Mirror Symmetry, $N=1$ Superpotentials and Tensionless
                  Strings on Calabi--Yau Four-Folds},
hep-th/9610162}

\lref\JMDN{J.~Minahan and D.~Nemeschansky,
{\it An N=2 Superconformal Fixed Point with $E_6$ Global
                  Symmetry}, hep-th/9608047;
{\it Superconformal Fixed Points with $E_n$ Global Symmetry},
hep-th/9610076}

\lref\WLNWa{W.~Lerche and N.P.~Warner, {\it Exceptional SW Geometry
from ALE Fibrations}, \hth/9608183.}
\lref\KLMVW{A.~Klemm, W.~Lerche, P.~Mayr, C.~Vafa and N.P.~Warner,
\nup 477 (1996) 746}
\lref\HT{C.\ Hull and P.\ Townsend, \nup438 (1995) 109,
hep-th/9410167.}

\lref\SW{N.\ Seiberg and E.\ Witten, \nup426(1994) 19;
\nup431(1994) 484}
\lref\FHSV{S. Ferrara, J. A. Harvey, A. Strominger and C. Vafa,
\plt361 (1995) 59}
\lref\Arn{See e.g., V.\ Arnold, A.\ Gusein-Zade and A.\
Varchenko,
{\it Singularities of Differentiable Maps I, II}, Birkh\"auser 1985.}
\lref\KKLMV{S.\ Kachru, A.\ Klemm, W.\ Lerche, P.\ Mayr and
C.\ Vafa, \nup459 (1996) 537}
\lref\kthree{A.\ Klemm, W.\ Lerche and P.\ Mayr, \plt357 (1995)
313;\br
P.\ Aspinwall and J.\ Louis, \plt369 (1996) 233}
\lref\KV{S.\ Kachru and C.\ Vafa, \nup450 (1995) 69}
\lref\ly{B.\
Lian and S.-T.\ Yau, \cmp176 (1996) 163.}
\lref\vg{C. Vafa and D.\ Goshal, \nup 453 (1995) 121.}
\lref\TMHS{
{T.\ Masuda and H.\ Suzuki,
 \nihil{Periods and Prepotential of $N=2$ $SU(2)$ Supersymmetric
Yang-Mills Theory with Massive Hypermultiplets,}
 \eprt{hep-th/9609066}.}}
\lref\KMV{A.\ Klemm, P.\ Mayr and C.\ Vafa,
 \nihil{BPS states of exceptional non-critical strings,}
 \eprt{hep-th/9607139}.}

\lref\ganorII{O.\ Ganor,
 \nihil{Toroidal Compactification of Heterotic 6D Non-Critical
Strings Down to Four Dimensions,}
 \eprt{hep-th/9608109}.}

\lref\witcom{E.\  Witten,
 \nihil{Some Comments on String Dynamics,}
 \eprt{hep-th/9507121}.}

\lref\swsix{N. Seiberg and E. Witten, \nup471 (1996) 121 }

\lref\ganor{
{O. \ Ganor,
\nihil{Six-dimensional Tensionless Strings in the Large N Limit,}
 \eprt{hep-th/9605201};}
{\nihil{Compactification of Tensionless String Theories.,}
 \eprt{hep-th/9607092}.}
}

\lref\rigid{A.\ Ceresole, R.\ D'Auria and S.\ Ferrara,
 \ex{On the Geometry of Moduli Space of Vacua
  in $N=2$ Supersymmetric Yang-Mills Theory,}
 Phys.\  Lett.\ {\bf B339} (1994) 71-76}

\lref\JDAH{J.\ Distler and A.\ Hanany,
 \nihil{(0,2) Noncritical Strings in Six Dimensions,}
 \eprt{hep-th/9611104}.}


\Title{\vbox{
\hbox{CERN-TH/96-326}
\hbox{USC-96/026}
\hbox{\tt hep-th/9612085}
}}{Non-Critical Strings, Del Pezzo Singularities}
\vskip-1cm
\centerline{{\titlefont and Seiberg--Witten Curves}}

\bigskip
\centerline{W. Lerche$^{a}$, P. Mayr$^{a}$ and N.P.~ Warner$^{b}$}
\bigskip
\bigskip\centerline{\it $^{a}$Theory Division, CERN, 1211 Geneva 23,
Switzerland}
\centerline{\it $^{b}$Physics Department, U.S.C., University Park,
Los Angeles, CA 90089}

\vskip 0in
\abstract{We study limits of four-dimensional
type II Calabi--Yau compactifications with vanishing four-cycle
singularities, which are dual to $\IT^2$ compactifications of the
six-dimensional non-critical string with $E_8$ symmetry. We define
proper sub-sectors of the full string theory, which can be
consistently
decoupled. In this way we obtain rigid effective theories that have
an intrinsically stringy BPS spectrum. Geometrically the moduli
spaces
correspond to special geometry of certain non-compact Calabi--Yau
spaces of an intriguing form. An equivalent description can be given
in terms of Seiberg-Witten curves, given by the elliptic simple
singularities together with a peculiar choice of meromorphic
differentials. We speculate that the
moduli spaces describe non-perturbative
non-critical string theories.
}

\Date{\vbox{\hbox{\sl {December 1996}}
\hbox{USC-96/026}
\hbox{CERN-TH/96-326}}}
\goodbreak

\parskip=4pt plus 15pt minus 1pt
\baselineskip=15pt plus 2pt minus 1pt

\newsec{Introduction}

The relation between singularities in supersymmetric string
compactifications on Calabi--Yau manifolds and field theories has
added much to the understanding of non-perturbative string theory in
the last two years. Starting from the interpretation of three-fold
singularities in terms of known field theories, the direction has
veered around in the meantime to the discussion of novel types of
theories to interpret various types of geometrical singularities.
Indeed one may expect to find in this way a whole zoo of
supersymmetric effective theories, and there is no reason why such
theories should always be interpretable in terms of conventional
field theories.

In particular, new theories have been discovered that involve
non-critical
strings in various dimensions
\refs{\witcom{,}\swsix{,}\eestring{,}\witm{,}\ganor{,}
\ganorI{,}\KMV{,}\ganorII{,}\witnew{,}\sp{,}\JDAH}. These
theories are still difficult to access directly and it is easier to
study them after further compactification. For example,
new interacting fixed points of five-dimensional gauge theories
\seibI\ have been related \doubref\ms\dkv\ to $S^1$ compactifications
of the non-critical string with global $E_8$ symmetry in six
dimensions. It is interesting to further compactify this theory on
another circle, and to investigate the properties of the resulting
$d=4$ \nex2 supersymmetric theory.

In this note we consider certain limits of four-dimensional
Calabi--Yau compactifications of type II strings, which are dual to
compactifications  of the six-dimensional non-critical $E_8$ string
on $\IT^2$.
Aspects of the resulting four-dimensional theory have been
discussed previously in \ganorII, and more recently, while
we were completing this paper, in \gms.
However, we will study different aspects of this theory, and, in
particular, our focus will be on a quite different, more stringy type
of moduli space. More specifically, we will obtain some new insights
in this moduli space by making use of the special geometry of
certain Calabi--Yau manifolds.

Recall that the non-critical string with global $E_8$ symmetry arises
in six dimensional Calabi--Yau compactifications of F-theory, when a
del Pezzo 4-cycle\foot{$\IB_n$ denotes the del Pezzo surface obtained
from blowing up $\IP^2$ at $0< n \leq 9$ points. For more details we
refer to \dkv.} $\IB_n$ shrinks to zero size \doubref\mv\witm. At
present, no simple representation of this six-dimensional theory in
its own is known, and its existence is based on a collection
of evidence rather than on a firm proof. If the six-dimensional
string
theory really exists, its effective theory should show up as a
``closed subsector'' of the full type II compactification.
Specifically, one should be able to separate off a sub-sector of the
BPS states of the full type II theory, which corresponds to
the embedded six-dimensional theory. This sub-sector should be
universal and should not depend on the details of the chosen
embedding three-fold.

We indeed find such a sub-theory, which we associate
to a certain non-compact
Calabi--Yau manifold as the underlying geometrical object.
This structure might hint at a more fundamental formulation
of the theory. A peculiar feature of this
sub-theory is that it involves the large complex structure limit of
the three-fold and thus is inherently stringy.

In section 2 we define consistent sub-moduli spaces of the
special geometry describing the effective type II theory, in
terms of monodromy and intersection properties. In section 3
we investigate two different sub-moduli spaces related to the tension
of the compactified non-critical string, for a specific embedding
three-fold, and discuss their physical properties.

In the rest of the paper, we then concentrate
only on one of these theories, namely on the one
whose moduli space intersects the large complex structure limit.
This moduli space corresponds, essentially, to the non-critical
string tension.
In section 4, we identify the underlying universal geometry
in terms of certain Landau-Ginzburg potentials, which involve a
coupling
of a $x^{-n}$ potential to the local geometry of del Pezzo
singularities.
The geometrical periods on the associated non-compact Calabi--Yau
space turn out to be closely related to those of certain elliptic
curves.
Specifically, we show in section 5  that the local three-fold
geometry can be captured by Seiberg-Witten curves with a
particular choice of meromorphic differentials. We then explicitly
evaluate in section 6 the periods near the singular
points in the moduli space. We verify that the theory is conformal
at the origin of
the moduli space, and reproduce from the SW formulation of the theory
its unusual stringy behaviour near infinity.
Finally, in section 7 we conclude with some more speculative remarks.

\nobreak

\newsec{Definition of the moduli space}

We want to extract the physics of the compactified
six-dimensional non-critical $E_8$ string (henceforth
referred to as ``NCS'') from the type II moduli space.
For this, it will be important to have a clear distinction
between the degrees of freedom that are part of the NCS
theory, and those that are ``superfluous'' states of the type II
theory.   The states of the  NCS will have a universal
description that is independent of the particular Calabi--Yau
manifold $\x$, whereas the superfluous states will depend
upon non-universal, global properties of $\x$.

Near a singular locus in the Calabi--Yau moduli space $\cx M (\x)$,
the theory is dominated by a subset $\cx H$ of the BPS states
of the full theory, with a typical mass scale $\Lambda$
much smaller than that of the rest of the
theory. An important characteristic of the subset $\cx H$ is the
behaviour of the theory away from the singularity. There
are, roughly speaking, two possibilities: either the
subset of states $\cx H$ does not extend to a globally well-defined
theory away from the particular scaling limit, so that it has to be
coupled to (a larger subset of) the rest of the theory; this
is, for example, what happens with asymptotically non-free field
theories. Or, more interestingly, the subset $\cx H$ itself provides
a globally well-defined theory also away from the singularity, as in
SYM theories with asymptotically free spectrum \SW.

The effective abelian $N=2$ gauge theory in four dimensions is
described by the holomorphic prepotential $\cx F$ of special
geometry, which itself is determined by the period integrals of $\x$.
The question of whether the truncated theory obtained in some scaling
limit is consistent and has a global extension, translates to the
question whether we can find a proper sub-moduli space $\cx M$ of the
full moduli space $\cx M(\x)$. This suggests a general definition of
``consistent sub-theories'' embedded in string theory, in terms of
{\it closed sub-monodromy problems}. That is, the full compactified
string theory is governed by a monodromy group $G$, and we look for a
subgroup $H \subset G$ that acts reducibly on the full set of BPS
states and closes on the subset $\cx H$ of states that are relevant
near the given singularity. If such a subgroup exists, we can
consistently throw away the remaining sectors of the theory and
define a new theory on the moduli space $\cx M_{H} \subset \cx
M(\x)$. One expects that these monodromy data can be associated to a
geometrical object, like the Riemann surface in SYM theories.

In fact, since it is known that the generic monodromy group
of a Calabi--Yau
compactification is generated by the monodromy on a generic
hyperplane in the (properly resolved) moduli space, we could start a
systematic search for such theories by classifying the (intersections
of) non-generic hyperplanes in $\cx M(\x)$.
Instead we will restrict ourselves to sub-problems which involve a
large complex structure point. In this case, it is simple to find
a necessary condition for the existence of a subgroup of the
monodromy
group. At a large complex structure point with maximal unipotent
monodromy, the Calabi--Yau periods, when
written in special coordinates $t_i$,  have the asymptotic
form:
\eqn\cyp{
\Pi \sim (1, \ \ t_k,\ \ \h c_{ijk}t^it^j+\dots,\ \  -{1\over 6}
c_{ijk}t^it^jt^k+\dots) \ ,}
where the $c_{ijk}$ are the triple intersections on
$\x$ and the ellipses denote subleading terms in the limit
$t_i \to \infty$ (which behave in a similar way).
It follows that a necessary condition for the existence
of a subset $S(H)$ of periods closed under the monodromy at infinity
is
that their intersection form with the rest of the periods vanishes.
Geometrically this means that we can find a
basis\foot{In some cases it may be useful to relax the condition
that the basis is integral,  in particular if the non-integrality
is related to elements not in $ S(H)$.}
of divisors in $ H_4(\x,\IZ)$, such that the subset $\cx S(H)$
does not
intersect with divisors that are not in $\cx S(H)$.

The above implies a rather interesting property of this
kind of sub-monodromy problems:
in contrast to the embeddings of SYM theories considered
previously \doubref\KKLMV\KLMVW,
these theories
inherit the instanton expansion - that is, the Gromov--Witten
invariants - from the Calabi--Yau manifold. In the special
sub-monodromy problem that we will
consider below, this is related
to the fact that the
underlying geometry is that of a non-compact Calabi--Yau
space of a very special form. This suggests an alternative
interpretation of the splitting condition on the intersection
form, namely in terms of
the existence of a well-defined, possibly non-compact
Calabi--Yau space that captures the relevant part of the
geometry.

There is also
the possibility of treating some of the Calabi--Yau moduli
as background fields rather than as dynamical quantities. In this
case the restriction on the intersections has to be imposed only
on the periods treated as moduli. An obvious example are
mass parameters in an effective field theory, which
originate as gauge masses in the full string theory.
\goodbreak
\subsubsec{The NCS as a sub-monodromy problem}\br
We now apply the above considerations to the elliptically
fibered three-fold $\xf$ with base $\IF_1$, which provides
the simplest embedding for the kind of theory we want to study.
Note, however, that the subsector corresponding to the NCS depends
only
on the local geometry, and that
the results we will obtain turn out to be
independent of the three-fold that we choose as an embedding.

F-theory on $\xf$ is dual to the heterotic string on $K3$ with
instanton embedding ($n_1$=11, $n_2$=13) in the two $E_8$ factors,
respectively. The vanishing 4-cycle singularity of the Calabi--Yau
compactification maps to the strong coupling singularity of the
heterotic dual. At this singularity, a non-critical string becomes
tensionless. It is the same string \doubref\witm\mv\ as the one
stretched
between the fivebrane and the ninebrane \eestring\ in M-theory on
$\IS^1/\IZ_2\times K3$, for a reason explained in \witnew.

The four-dimensional $N=2$ supersymmetric
theory, obtained by compactification on
a further $\IT^2$, is dual to the type IIA string
on $\xf$. There are two
Calabi--Yau phases, which are related by a flop;
we denote them by Phase $I$ and and Phase $II$, respectively.
The  $h^{1,1}=3$ vector multiplet fields $t_E,\ t_F$ and  $t_B$
parametrize in Phase $I$ the volume of a curve in the fiber
and the two classes of $\IF_1$, respectively. The base
of $\IF_1$, with volume $t_B$, is also the base of a K3
fibration and is therefore related to the heterotic dilaton.
Figure 1
shows a real slice $it_i \in \IR$ through $\cx M(\xf)$, and
displays the singularities associated to
the flop and the vanishing 4-cycle, respectively.

The leading pieces of the prepotentials
$\cx F_I$, $\cx F_{II}$, which in particular
summarize the intersection properties of divisors in $\xf$,
are given in Appendix A. A characteristic
split of the intersection form, indicating a
possible closed sub-monodromy problem which intersects the large
complex structure point, can be observed in Phase $II$.
More precisely, in a basis, where the scalar
fields in the vector multiplets are
given by $t_S\equiv\tee+\teb,\ \tfp\equiv\tef+\teb$, and  $\tee$,
there are three periods which depend classically only on one
of the three fields.

\goodbreak\midinsert
\centerline{\epsfxsize 4.truein\epsfbox{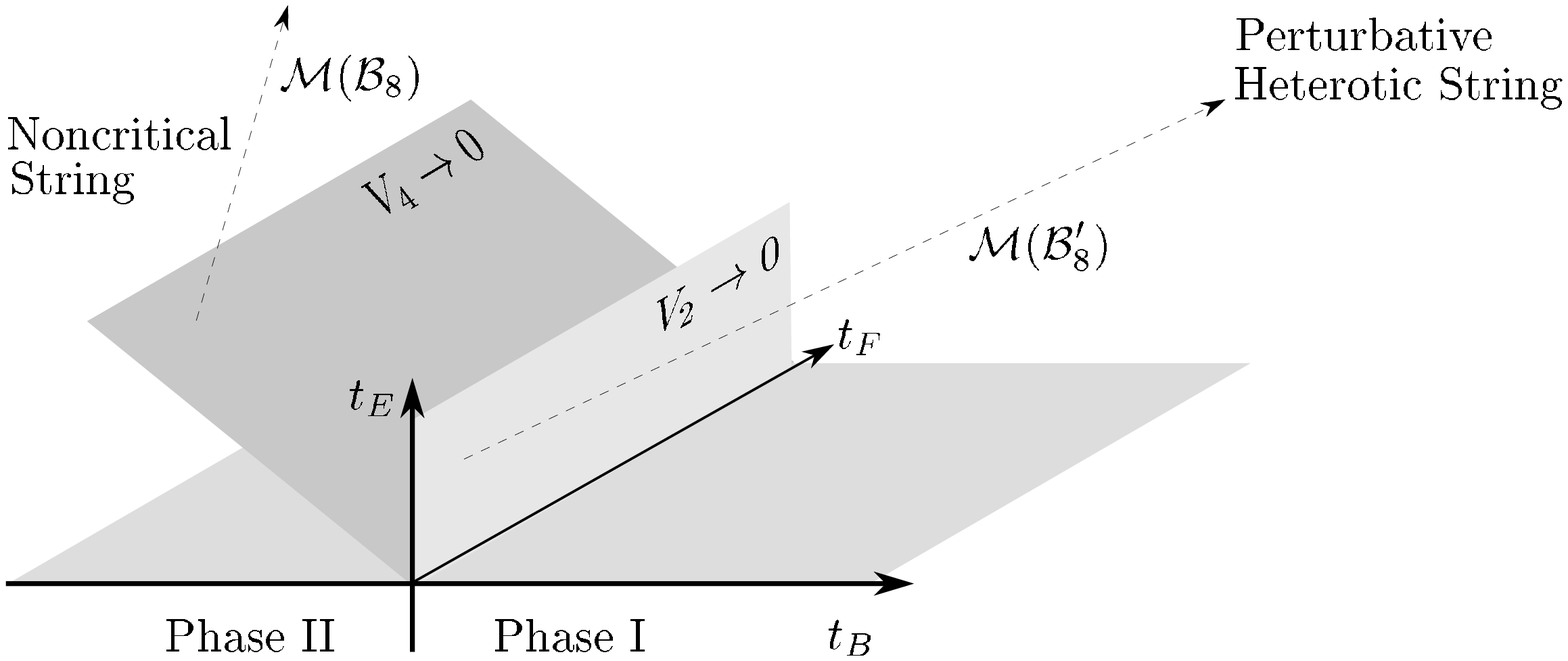}}\leftskip 2pc
\rightskip 2pc\noindent{\ninepoint\sl \baselineskip=8pt {\bf Fig.1}:
Real slice through the K\"ahler moduli space of $\xf$. The planes
$V_2\to 0$ and $V_4\to 0$ denote the faces of the K\"ahler cone where
a 2-cycle (flop) and a 4-cycle (del Pezzo) shrink, respectively. We
refer to the phases with $t_B > 0$ ($t_B < 0$) as Phase $I$ ($II$).
We have also sketched the moduli spaces of the effective theories
$\cB_8$, $\cB_8'$ of the NCS that we will consider below. Far out at
infinity in Phase $I$ is the region of the perturbative heterotic
string. On the other hand, Phase $II$ is not a $K3$ fibration and
thus not related to the perturbative heterotic string.}
\endinsert

Continuation beyond the flop
reveals that this splitting property is no longer present
in Phase I.
For a globally consistent definition of this system
beyond the flop, we have to include one further period as a
background parameter, namely the volume of the
elliptic fiber $t_E$. So depending on whether we include $t_E$,
or not, we have two subsets of periods that define appropriate
sub-monodromy problems:
\eqn\sbp{
\tilde{\Pi}_i = (1,\ t_S,\ \partial_{t_S} \cx F) \hskip 2cm
\tilde{\Pi}^\prime_i = (1,\ t_S,\ \partial_{t_S} \cx F,t_E) \ .
}
In the following we will mainly focus on the three periods
$\tilde{\Pi}_i$, which correspond to $0,2,4$ cycles in the
three-fold,
respectively; the 4-cycle is the del Pezzo $\IB_8$ in question.
It is gratifying to observe that the single K\"ahler
coordinate that is treated as a modulus in $\tilde\Pi_i$,
namely $t_S$, is precisely the one which parametrizes the
volume of the 4-cycle.\foot{More generally, one could consider
higher dimensional moduli spaces by including additional
homology classes of the del Pezzo surface, corresponding
to non-zero Wilson lines of the heterotic theory.}

For a complete definition of the theory
we have still to choose values for the other moduli,
at which we vary $t_S$. In fact,
different choices will lead to rather different
physics. In particular, the BPS spectrum at the singularities
of the moduli space $\cx M_S$ will depend which set of the
Calabi--Yau singularities is  hit by this moduli space.

\newsec{Effective theories of the NCS}
\def\ppI{(A.1)}
To relate the physics of the embedded theory, defined by the
periods \sbp, to the NCS, we have to find the relation between the
Calabi--Yau moduli and the fields of the effective theory of the NCS.
This can be done by extending the five--dimensional analysis of
\doubref\ganorI\KMV. In the five--dimensional compactification, the
radius $R_6$ of $\IS^1$ is inversely proportional to the size of the
elliptic fiber, $t_E$. Moreover, the tension of the magnetically
charged string is proportional to the volume $V_4$ of the 4-cycle
$\IB_n$.
This volume can be parametrized by the K\"ahler coordinate $t_S$,
which
measures the volume of 2-cycles within $\IB_n$. More precisely
one has: $V_4 \sim t_S^2$.

For a further $\IS^1$ compactification of the
five-dimensional theory to four dimensions, the relation between the
coordinates is then given by $t_i^{(4)}=iR_5 t_i^{(5)}$ \gun.
This leads to the approximate identifications:
\eqn\ncsc{
 t_S =: \phit=i \phi R_6 R_5, \quad \ \tee =: \cx U = i
{R_5 \over R_6}, \quad
\tfp =: \cx T = i \phi_F R_6 R_5 + \cx O(R_6^0)\ ,
}
where $\phi$ ($\phi_F$) is the tension of the non-critical
(fundamental)
string\foot{We set $\phi_F=1$ if not denoted otherwise.}.

The precise identifications, which take a non-diagonal metric of the
torus
and the $B$-field into account,  can be inferred from the map to the
moduli of the heterotic dual in the weakly coupled regime.
These moduli are the dilaton $S$, and the K\"ahler and
complex structure of the torus,
i.e. $T=B+iR_5R_6$, $U=  e^{i \alpha}R_5/R_6 $.
Comparing the prepotential $\cx F_I$ \ppI\ of the K3
fibered Phase $I$ with that of the heterotic string, where
$\cx F_{het}=STU+\cx O(e^{2\pi i S})$, one obtains\foot{The constants
$a$ and $b$ remain undetermined in perturbation theory.}
$\tef = T-U,\ \teb = S+a\ T+b\ U,\ \tee = U$ (here we
have assumed that $Im(T) > Im(U)$).

As mentioned before, the effective $N=2$ $U(1)$ theory
is determined after fixing the values of $\cx U$ and $\cx T$.
The theory obtained in this way can be suitably characterized
by the surviving BPS spectrum. For this, let us
first reconsider the five-dimensional
theory, which contains the relevant features of the
four-dimensional case.

Point-like BPS states in five dimensions have
masses determined by the central charge \fer
\def\tf{t^{(5)}}
\eqn\bpsf{\eqalign{
 Z=\sum_in_i\tf_i
&=
{1\over \cx V_5^{1/3}}
( l \cdot \phi R_6 + p \cdot {1 \over R_6}  +m \cdot  (\phi R_6
-2{1  \over R_6}+ R_6)) \ ,
}}
where the $\tf_i$ are subject to the constraint
$\cx V_5(\tf_i)=1$, with $\cx V_5(v)=c_{ijk}v_iv_jv_k$
and $c_{ijk}$ the intersection numbers \gun.

Phase I (II) corresponds to the region $\phi R_6 >  {1\over R_6}
\ (\phi R_6<  {1\over R_6})$.
As shown in Fig.\ 1,
far out in the K\"ahler cone of Phase I is the regime of the
perturbative
heterotic  theory, and the light states are the fundamental string
windings
and momenta with quantum numbers $q\equiv (l,p,m)=(l,p,-l)$. As one
increases the
tension $\phi$, one hits the boundary between phases $I$ and $II$.
Associated to the vanishing 2-cycle is a massless hypermultiplet
\witm\ with
winding/momentum $q=(l,p,m)=(1,-1,0)$ \KMV.
In fact, in this limit there is also a net number of 480
fundamental string momentum modes that are equally relevant.
They correspond to rational curves in the fiber and their number has
a
simple interpretation: T-duality in the fiber transforms them
to 0-branes, whose moduli space is the Calabi--Yau $\xf$
itself - with Euler number $\chi=-480$.

Geometrically, the 4-cycle which is shrunk at
the singularity $\phit=0$
is still of type $\IB_9$ in Phase $I$ and becomes
type $\IB_8$ in Phase $II$.
It is thus  clear
that there will be two different classes of effective theories, whose
moduli spaces intersect at the singularity $\phit=0$:
for any finite value of
$\cx U$, increasing the string tension $\phi$ will finally let us
hit the wall $\phi R_6 = {1\over R_6}$, where an additional
hypermultiplet becomes massless and changes the asymptotic
behaviour of the  theory. In fact, the same will happen when
starting from the $\IB_n$ type of singularity, where one hits
successively the walls corresponding to massless hypermultiplets,
as is clear from the alternative description in terms of D4-branes
\ms\dkv.
In physical terms one ends up with a matter content $N_f=8$ of
an $SU(2)$ gauge theory, whose Coulomb branch is parametrized
by $\phit$. This theory has a trivial dependence
on the Coulomb modulus, that is,
a constant effective coupling \seibI .
We will refer to this kind of theory, which starts from the
$\IB_n$ singularity and ends up in the $\IB_9$ phase, as $\Bp_n$
\foot{These theories have been considered in \gms.}.

On the other hand, if one wants to keep the theory in the phase with
a 4-cycle of type $\IB_8$, one has to send $\cx U$ to infinity first,
such that the wall of the flop is shifted away by an infinite
distance. We will refer to this type of sub-theory as $\cx B_n$.
It is this kind of theories that we will mainly focus on in this
paper.

The situation is similar after further compactification to
four dimensions: for any finite value of the modulus $\cx U$, the
four-dimensional theory reaches phase $I$ for large values of the
string tension, and transmutes to the $\cx B_9$ type of theory.
We will denote the resulting theories
by $\cx B_n$ and  $\cx B_n'$ as well.
The locations of the moduli spaces of the theories $\cx B_8$ and
$\Bp_8$ in $\cx M(\x)$ are indicated in Fig. 1.
\goodbreak
\subsubsec{The effective theory $\cx B_8^\prime$}\br
This is the theory with a trivial asymptotic behaviour.
The periods dual to $\phit$, obtained from \ppI, are:
\eqn\ppII{\eqalign{
{\rm Phase}
\ I:\qquad &  \phitd = -\h \cx U^2 + \phit\ \cx U -\h \cx U - \h +
\partial_
{\phit} \cx F_I^{inst} \ , \cr
{\rm Phase}
\ II:\qquad & \phitd = \h  \phit^2 - \h \phit -{5 \over 12}+
\partial_
{\phit} \cx F_I^{inst} \ .
}}
Restricting to the quadratic terms in \ppII\ one recovers the
five-dimensional expressions \doubref\ms\dkv. The
four-dimensional BPS mass formula reads in special coordinates \cerI:
\eqn\bpsI{
Z=e^{K/2}(1\cdot n_0+t_i\cdot n_i-\cx F_0 \cdot m_0 - \cx F_i \cdot
m_i) \ ,
}
where $K$ is the K\"ahler potential, $\cx F_i = \partial_{t_i} \cx F$
and
$\cx F_0 = 2 \cx F - t_i \cx F_i$. Moreover, $n_A$ and $m_A$,
$A=0\dots 3$,
are integer charges.
Restricting to the periods of the second sub-problem in \sbp,
$\tilde\Pi_i' = (1,t_S,\partial_{t_S}\cx F,t_E) \equiv
(1,\phit,\phitd,\cx U)$,
is justified if all other periods are much larger.
To achieve this, we have still to decouple part of the
type IIA string states. The only modulus
that involves explicitly
the fundamental string tension $\phi_F$, is $\cx T$:
$$
Im(\cx T) = \phi R_5 R_6 \sin \alpha -2 R_5/R_6 \sin \alpha
+ \phi_F R_5R_6 \sin\alpha  \ .
$$
Thus, in order to decouple the fundamental string states,
we send $\phi_F \to \infty$ and obtain:
\eqn\ccII{\eqalign{
Z &\ \ =\ R_5e^{K/2}\big
({n_0\over R_5} + n_2\phi R_6 + {n_3\over R_6} - m_2 \phitd\big)\ .
}}
{}From \ccII\ and $\phitd \sim \phit R_5 + \cx O(\phit^0)$ we see
that
the states with masses $m=|Z|$ can be interpreted as
the winding and momentum states of the NCS on the torus. Furthermore,
as can be seen from \ppII,
the monodromy at infinity $\phit \to \phit +1 $ acts on the periods
as
$$
(1,\phit,\phitd,\cx U) \quad \to \quad (1,\phit+1,\phitd +\cx U, \cx
U)\ .
$$
It acts trivially on the $U(1)$ gauge coupling,
$\tau(\phit)  = \partial_{\phit}\phitd$, as has been observed
also in \gms.
\goodbreak
\subsubsec{The effective theory  $\cx B_8$ }\br
The behaviour of the $\cx B_8$ theory turns out to be rather
different.
As mentioned before, if we send $\cx U \to \infty$ first, we stay
always
in Phase $II$. There is no need to include the period $\cx U$,
because
the charged states associated to it have become infinitely massive.

The gauge
coupling behaves like $\tau(\phit)\sim \phit$, as is familiar
from the large complex structure limit; it is typical
for a decompactification limit.
{}From \ppII\ and \bpsI\ one finds that for large
$\phit$, the BPS spectrum contains states with masses
\eqn\bpsz{
m \sim \qquad \phit\ ,\quad 1\ , \quad
{1\over \phit}
}
up to multiplication with a common mass scale.
In particular, since $\phit=b+i R_5R_6\phi$, these masses
depend only on the $b$ field and the tension times a volume
factor, which in turn can be reabsorbed in the tension. This is
very much like in a ``pure'' uncompactified string theory,
where there are no relevant geometrical parameters to be varied.

{}From the type IIA perspective the above states correspond to
4, 2 and 0 brane states, the latter ones being the relevant
states for large $\phit$, uncharged under the $U(1)$ gauge
symmetry.
The states whose mass depends inversely proportional
to the volume can be interpreted
as Kaluza--Klein modes. The states with masses $\sim \phit, \sim 1$
(up to a prefactor $\phit$) are the
ones which become massless at $\phit=0$ (despite quantum
corrections).
There is one further massless state for a value 1/2 of the
$b$ field, namely a 0-4 brane bound state, which leads to
a conifold singularity, as we will see later on.

\subsubsec{Gauge couplings in five and four dimensions}\br
A crucial difference between the five--dimensional and
four--dimensional fixed point theories is the value of the $U(1)$
gauge coupling, $\tau=\partial_{\phit} \phitd$ (cf., \ppII) at the
critical point $ \phit\sim \ep \to 0$: it is infinity in five \seibI,
but should be at a fixed point of the modular group in four
dimensions \JMDN.

Geometrically, infinite coupling in five dimensions is the statement
that if the 2-cycle volume goes to zero as $V_2=\phit \sim \ep$,
the 4-cycle volume vanishes as $\phit^2 \sim \ep^2$.
This follows from the quadratic terms in \ppII:
\eqn\gc{
{1\over g_0^2} \sim \partial_{\phit} \phitd \sim \ep\ .
}
In four dimensions, quantum corrections do not lead to a finite
volume for neither the 4-cycle nor the 2-cycle \sp, so that cannot be
the explanation for the finite value of the coupling constant.
However, from \gc\ it is clear that it is enough that the instanton
corrections change the scaling from $V_4 \sim V_2^2$ to $V_4 \sim
V_2$. We will see below in section 6 that this is precisely what
happens, that is, $V_4 \sim \tau \ V_2$ with $\tau_0$ the
gauge coupling at the fixed point.

This has an interesting consequence which could have been anticipated
on physical grounds. As explained in \KMV, there are states in the
four-dimensional type IIA theory with masses $\sim V_2,\ \sim V_4,\
\sim V_4^{\h}$ that arise from wrapping the D2, D4 and 5 brane,
respectively. Would we still have $V_4 \sim V_2^2$ as in five
dimensions, the most relevant states would be the wrapped D4 branes
and this would lead to a theory with mutually local ``magnetic''
charges. This would be in contradiction with the experience that
conformal fixed points always involve states with mutually non-local
charges \witx. However, if $V_4 \sim V_2$, the D2 brane state is as
relevant as the one from the D4 brane, implying the presence of
mutually
non-local massless BPS states.

\newsec{Intrinsic geometrical formulation of the theories $\cx B_n$}

We will now consider in some more detail the four-dimensional
theories $\cx B_n$ ($n=6,7,8$) as closed sub-theories by
themselves. In particular, we will formulate these
theories independently of an embedding in a
``larger'' Calabi--Yau manifold.

Let us first ask about the geometrical object behind the
sub-monodromy problem defined and described above. It is a
remarkable fact that we can associate Gromov-Witten invariants with a
sub-monodromy problem that depends only on a single modulus $t_S$
(associated to the volume of 2-cycles in
the del Pezzo surface $\IB_n$). One
may wonder about why this separation of the del Pezzo surface from
the whole Calabi--Yau three-fold works, since the del Pezzo has a
non-vanishing first Chern class, $c_1 \neq 0$; one thus expects a
non-trivial coupling to two-dimensional gravity. Recall therefore the
description of the conifold singularity in ref. \vg. Although the
conifold geometry is locally not Ricci flat, it can be described by a
Landau-Ginzburg model which combines the description of the $\ch=1$
string in terms of the WZW theory $SU(2)$ at level $k=-3$ \chone,
with an ALE space with $A_1$ singularity:
\eqn\cof{
W_{CF} = {1 \over x} + y_1^2 + y_2^2 + y_3^2 + y_4^2\ .
}
This Landau-Ginzburg potential is related to a hypersurface
$\IP^{-2,1,1,1,1}[2]$ and satisfies the Calabi--Yau condition with
$\ch=3$. ADE type of generalizations of this Landau-Ginzburg
potential, which involve the coupling of $1/x^\ell$ potentials to ALE
spaces, have been considered in \ovI.

In the present case, the local geometry of the singularity is
described by the vanishing del Pezzo 4-cycle $\IB_n$ instead of an
ALE singularity. We are thus led to consider the following
Landau-Ginzburg potentials:
\eqn\delPezz{\eqalign{
  \IP^{3,2,1,1,-1}[6] &:\ \ W_{E_8}\ =
{1\over x^6} + y^2+w^3+z_1^6+z_2^6 -\psi\, xyz_1z_2w\cr
 \IP^{2,1,1,1,-1}[4] &:\ \  W_{E_7}\ =
{1\over x^4}+ y^2+w^4+z_1^4+z_2^4-\psi\, xyz_1z_2w\cr
  \IP^{1,1,1,1,-1}[3] &:\ \  W_{E_6}\ =
{1\over x^3}+ y^3+w^3+z_1^3+z_2^3-\psi\, xyz_1z_2w\ \cr
  \IP^{1,1,1,1,1,-1}[2,2] &:\ \
W^1_{E_5}\ =
{1\over x^2}+z_1^2+z_2^2+z_3^2 -\psi\, xz_3z_4z_5\ ,\cr
&\hskip 5.5mm W^2_{E_5}\ = z_3^2+z_4^2 -\psi\, z_1z_2\ ,
}}
which describe non-compact Calabi-Yau spaces.
The power $\ell$ of $x$ is determined by the Calabi--Yau condition,
$\ch=3$. We have also indicated the weights of the non-compact
Calabi--Yau spaces, as well as the canonical hyperplane perturbation
$\psi$ which deforms away from the Landau-Ginzburg point.
Specifically, $\psi=\infty$ corresponds to the large
complex structure limit, while the symmetric point $\psi=0$ gives the
zero size limit of the del Pezzo 4-cycle. A natural coordinate on the
moduli space $\cM(\cB_n)$ is given by $\ca=\psi^{-\ell}$, where
$\ca=(e^{2\pi i t_S}+$instanton corrections) and $\ell=6,4,3,4$
for the three cases, respectively.

The point is that the period integrals over the
corresponding holomorphic 3-forms
\eqn\intOmega{
\tilde\Pi_i(\ca) = \int_{\Gamma_i} \Omega ~=~{\psi}\int_{\Gamma_i}
 {x\,dw\,dy\,dz_1 d z_2\over  W_{E_n}(\ca)}}
indeed reproduce the periods in \sbp\ that govern our
sub-monodromy problem. Here,
${\Gamma_i}$, $i=1,\dots,3$ denotes a basis of 3-cycles on the
surface $W_{E_n}=0$. These 3-cycles are related via mirror symmetry
to $0,2,4$-cycles, where the 4-cycle is the del Pezzo surface itself.
More concretely, the periods $\tilde\Pi_i$ are solutions of the
following third-order Picard-Fuchs operators associated with
the three-folds \delPezz:
\eqn\dopell{
\cx L^{(n)} = \cx L_{ell}^{(n)}\cdot \theta \ ,\qquad\ \
\theta\ \equiv\ \ca \,\Coeff\partial{\partial \ca}\ .
}
These form certain degenerate systems of generalized hypergeometric
type ${}_3F_2$, and are known \KMV\
to govern the relevant periods of the del Pezzo surfaces.
Above, the operators $\cx L_{ell}^{(n)}$, defined by
\eqn\pfI{\eqalign{
\cx L_{ell}^{(8)}&=\theta^2-12\ca\ (6\theta+5)(6\theta+1)\ ,
\qquad (\coeff16,\coeff56,1) \cr
\cx L_{ell}^{(7)}&=\theta^2-\ 4\ca\ (4\theta+3)(4\theta+1)\ ,
\qquad (\coeff14,\coeff34,1)   \cr
\cx L_{ell}^{(6)}&=\theta^2-\ 3\ca\ (3\theta+2)(3\theta+1)\ ,
\qquad (\coeff13,\coeff23,1)  \  \cr
\cx L_{ell}^{(5)}&=\theta^2-\ 4\ca\ (2\theta+1)(2\theta+1)\ ,
\qquad (\coeff12,\coeff12,1)  \ , \cr
}}
are the Picard-Fuchs operators \ly\ of the following
elliptic curves:
\eqn\tori{\eqalign{
\IP^{1,2,3}[6] &:\ \  \ P_{E_8}\ =\ y^2+w^3 + z ^6 - \psi\, ywz\ ,
\cr
\IP^{1,1,2}[4] &:\ \  \ P_{E_7}\ =\ y^2+w^4 + z ^4 - \psi\, ywz\ ,
\cr
\IP^{1,1,1}[3] &:\ \  \ P_{E_6}\ =\ y^3+w^3 + z ^3 - \psi\, ywz\ ,
\cr
\IP^{1,1,1,1}[2,2] &:\ \  \ P^1_{E_5}\ =\ y^2+w^2 - \psi\, v z \ , \
P^2_{E_5}\ =\ v^2+z^2 - \psi\, yw \ .
}}
They define ordinary
hypergeometric systems ${}_2F_1$ of the types $(\a,\b,\gamma)$
indicated on the r.h.s. of eq. \pfI. Therefore, the del Pezzo periods
are related to the
usual periods $(\varpi(\ca), \varpi_D(\ca))=\psi\!\int\! dw/y$ of the
tori \tori\ by
\eqn\thetamap{
\theta\cdot \tilde\Pi_i \ \equiv\ \theta\cdot
(1,\ \phit(\ca),\ \phitd(\ca)) \  =\
(\,0,\varpi(\ca),\,\varpi_D(\ca)\,)\ .
}

Interestingly one 
can treat the conifold singularity \cof\ in a similar
way, by adding the hyperplane perturbation: $W_{CF} = {1 \over x} +
y_1^2 + y_2^2 + y_3^2 + y_4^2- \psi\,xy_1y_2y_3y_4$. This leads to a
hypergeometric system of type $(\coeff12,\coeff12,1)$, similar as for
$E_5$, up to a subtle minus sign $\ca \to -\ca$
related to a half-integer shift of the B-field.
\goodbreak
\subsubsec{Topological properties, singularities and Gromov--Witten
invariants}\br
The Calabi--Yau spaces \delPezz\ turn out to have quite
interesting properties, suggesting a canonical structure also
for the cases $\cx B_n,\ n<5$.

The topological properties can be summarized in terms of the
triple self-intersections, the Euler number and an
integral over the second Chern class, $c_2$:
\eqn\topdat{\eqalign{
\int_X J\wedge J\wedge J &= n-9\ ,\cr \int_X J \wedge c_2 &= -12+2\
(n-9)\ ,\cr
\int c_3 &=\ \chi \ = \  - 2\ h(E_n)\ ,
}}
In particular, the Euler number is precisely twice the dual Coxeter
number
$h(E_n)$ of $E_n$! This agrees
nicely with the physical expectation about the moduli space of the
six-dimensional NCS, which involves the $U(1)$ gauge field and the
hypermultiplets of the instanton moduli space (there is in addition
one universal hypermultiplet which decouples together with gravity).
Although we have derived these formulas from \delPezz\
only for the values $n=8,7,6,5$,
it is clearly suggestive that they
should apply also for the other values of $n$.
In fact, the triple intersections agree with the
self-intersections of $c_1\in H_2(\IB_n,\ZZ)$ in $\IB_n$
(and with the self-intersections of the elliptic curves
\tori\ \SaitoES\ as well). These facts further suggest that we really
capture with the three-folds \delPezz\ the intrinsic information of
the
NCS associated with the del Pezzo surfaces.

There are three singular points with non-trivial monodromies
in the moduli spaces $\cx M(\cB_n)$, namely
i) the large complex structure point $\ca=0$, ii) the point $\ca =
\infty$,
where the 4-cycle volume vanishes and iii) a conifold point at
{$1+d_nc=0$}, where $d_n=432,\ 64,\ 27,\ 16,\ -16$
for $n=8,7,6,5,CF$, respectively.

Let us comment on the nature of this conifold point. Firstly, it does
not correspond to the conifold point that occurs in
$\cB_n'$ at the flops, e.g. $t_B=0$ for $\cx B_8^\prime$. As
indicated in Fig.1, the latter conifold point is not hit by
the moduli space $\cM(\cB_8)$. Secondly, for the $\cx B_n$ theories,
$1+d_nc=0$ requires negative values of $c$ and specifically
a half-integer value of the $b$ field. Therefore this singularity
is absent in the Lorentz invariant five and six dimensional
theories. On the other hand, $c$ is positive at the conifold
point of the ``conifold'' theory \cof, as it should be for
consistency.

At $c=-d_n^{-1}$ there is a simple pole in the Yukawa coupling
\eqn\yukII{
w(\ca) = {1\over {\ca}^3 (1+d_n \ca)} \ ,
}
which, similar as in the compact case, obeys
a linear differential equation,
$$
(3+4\ d_n\ \ca)\ w(\ca) + \ca \ (1+  d_n\ \ca)\  w^\prime(\ca) = 0 \
{}.
$$
The Gromov--Witten invariants can be obtained from transforming
\yukII\ to special coordinates centered
at the large complex structure point $\ca=0$. One finds
the following instanton expansions for the Yukawa couplings
$C_{\phit\phit\phit}\equiv\partial^2_{\phit}\phitd$:
\def\XYZ#1{{\scriptstyle{{#1^3  q^{#1}\over 1-q^{#1} } }}}
\eqn\yuk{\eqalign{
E_8 &:
-1+252\XYZ1-9252\XYZ2+848628\XYZ3-114265008\XYZ4+\dots \ , \cr
E_7 &:
-2+56\XYZ1-272\XYZ2+3240\XYZ3-58432\XYZ4+1303840\XYZ5+\dots \ , \cr
E_6 &:
-3+27\XYZ1-54\XYZ2+243\XYZ3-1728\XYZ4+15255\XYZ5+\dots \ , \cr
E_5 &:
-4+16\XYZ1-20\XYZ2+48\XYZ3-192\XYZ4+960\XYZ5 +\dots \ , \cr
CF &:
\ \ 4+16\XYZ1+16\XYZ2+48\XYZ3+192\XYZ4+960\XYZ5+\dots \ , \cr
}}
where $q\equiv e^{2\pi i \phit}$. The instanton numbers agree
with those calculated in a embedding Calabi--Yau for $n=8,7,6,5$
\KMV.

Interestingly, the expansion for the
conifold theory \cof\
is related to that of $\cx B_5$ by a simple change of
sign of $q$ in the Yukawa coupling, which corresponds to
a half-integer shift of the $b$ field, as mentioned before.
In particular, the non-compact Calabi--Yau space associated to $E_5$
and
the conifold share the functional dependence of the periods
on the modulus. The only difference is in the relative shift of the
integral symplectic bases.
\goodbreak
\subsubsec{Index of the singularities and elliptic curves}\br
As explained in \cvI, an important
quantity characterizing the massless spectrum at the
singularities is the singular behaviour of the topological amplitude
$F^{top}_1$, defined in \bcov. In the present case, the
ansatz for $F^{top}_1$ is particularly simple:
\eqn\ftop{
F^{top}_1 = \log \left[ \ca^a (1+d_n\ \ca)^b {d \ca \over dt}\right]
\ ,
}
where $a,b$ are constants. These can be determined by comparison
with the expression for $F_1^{top}$ in terms of
world-sheet instantons of genus zero and one\foot{
For the numbers $n^e_k$ of elliptic curves, see the Appendix.}:
\eqn\ftopII{{F_1}^{top}=-{2 \pi i \over 12}
t\!\int\!c_2 \wedge J-\sum_{k} \left[2 n^{e}_k\log(\eta(
q^{k}))+ {1\over 6} n^r_k
\log(1-q^{k})\right]+{\rm const.
}}
The massless spectrum at a singularity $\ca=\ca_0$ is then given by
the coefficient $\beta$ of the logarithm, $F^{top}_1 \sim -\beta/6\
\log(\ca-\ca_0)$,
which counts the net number of massless vector hypermultiplets minus
hypermultiplets, $\beta=n_H-n_V$.

At the conifold singularity, the above calculation yields
$\beta=1$, as
expected on general grounds. This singularity is therefore indeed due
to
one additional massless hypermultiplet.

More interesting is the behaviour at the point $\ca=\infty$,
where the 4-cycle vanishes. Taking carefully into account
the contributions from the three factors in \ftop, namely
a universal contribution $b=-{1\over 6}$
from the conifold discriminant, a factor $({7\over 6},
{5\over 4},{4 \over 3},{3\over 2})$ from the determinant
and $a=(-{1\over 6},-{1\over 3},-{1\over 2},-{2\over 3})$,
one finds
$$
\beta=(30,18,12,8) =  h(E_n)
$$
for $\cx B_n, \ n=8,7,6,5$,
respectively, where $h(E_n)$ is the dual Coxeter number of
the global symmetry group $E_n$.
The net massless spectrum at the vanishing 4-cycle singularity is
therefore that of $h(E_n)$ extra massless hypermultiplets. This is
in agreement with the change of hodge numbers proposed in
\mv\ in the context of Calabi--Yau transitions.

In view of the bad convergence of the instanton expansion \yuk\ at
the
conformal point $\ca=\infty$, this is quite remarkable.
In particular, the number of massless BPS states arising from
wrappings
of vanishing 2-cycles is unbounded, indicating a breakdown
of the description of the theory in terms of these degrees of
freedom.
The index $\beta$ sums up the contribution of the infinite number of
terms
to a finite result, indicating that the individual terms still make
sense
when appropriately regularized.
\goodbreak
\subsubsec{Comparison with SYM theories with matter}\br
Note that the relations \thetamap\ are very reminiscent of
Seiberg-Witten theory with extra matter \SW: derivatives of $\phit
\sim a,\phitd \sim a_D$ give torus periods. Moreover, the constant
period, whose effects reflect Kaluza-Klein excitations in the present
context (cf. \bpsz), corresponds to a bare mass. Indeed, the
$E_{6,7,8}$ theories can formally be associated to the $SU(2)$
theories with $N_f=1,2,3$, respectively. That is, when tuning the
mass parameters appropriately, one can check that the corresponding
SW curves become equivalent (up to isogeny) to the $E_{6,7,8}$ tori
\tori; this is related to observations made in \TMHS.
This implies that the effective $U(1)$ gauge couplings
\eqn\taudef{
\tau(\ca)\ \equiv\ {\del^2\over\del\phit^2} \cF(\phit)\ =\
{\del\over\del\phit}\phitd(\phit)\ =\
{\varpi_D(\ca)\over\varpi(\ca)}\ ,
}
are, when viewed as functions of $\ca$, the same as the field theory
couplings (e.g.., for small $\ca$, $\tau\sim \log(\ca)$).

However, the main difference from this viewpoint between the
Yang-Mills theories and the $\cB_n$ models is that the
``$\theta$-operators'', which map $\phit,\phitd$ to the torus
periods, are different. That is, $\theta=\ca{\del\over\del\ca}$ is a
logarithmic derivative and not an ordinary derivative. As a
consequence, the couplings $\tau(\phit)$ (and thus the prepotentials
$\cF(\phit)$) as functions of $\phit$ are different as compared to
ordinary field theories, and this implies that the theories $\cB_n$
describe completely different physics.

Nevertheless the formal similarity suggests that the relevant
geometry of the del Pezzo 4-cycle, and its moduli space $\cM(\cB_n)$,
can be captured by SW curves given by the tori \tori. For this to
work, however, suitable meromorphic differentials $\lambda$ should
exist on these curves that give rise to the periods $\tilde\Pi_i$.
We will verify in the next section that such meromorphic
forms can indeed be obtained directly from the 3-forms, $\Omega$.

\newsec{Geometric reduction to elliptic curves}

We first consider the surface $W_{E_8}=0$ in weighted $\IP^4$ for the
$E_8$ Landau-Ginzburg potential \delPezz. After re-defining $x$ and
$y$ by some trivial phases, this surface can be written:
\eqn\Eeight{y^2 ~=~ w^3 ~+~ z_1^6 ~+~ z_2^6  ~-~
{1 \over x^6} ~-~ \psi\, w x y z_1 z_2 \ .}
Passing to the patch  $x = 1$, and integrating \intOmega\
with respect to $y$ around a contour about the surface $W_{E_8} = 0$
yields:
\eqn\omreduce{\int \Omega ~=~ \int {\psi~dw~dz_1~dz_2\over  r} \ ,}
where
\eqn\denom{r ~=~ {\partial W_{E_8} \over \partial y} ~=~ 2~\sqrt{w^3
{}~+~
z_1^6 ~+~  z_2^6 ~-~ 1 ~+~ \coeff{1}{4} (\psi w z_1 z_2)^2 }\ .}

We now wish to integrate $\Omega$ over some very particular
$3$-cycles in the surface \Eeight.  To exhibit these cycles we
follow a simple variation of the approach of \refs{\KLMVW,\WLNWa,
\JMDN}.  First we slice the surface by the simple quadrics,
$w = \alpha z_1^2$, and use $\alpha$ and $z_1$ as new independent
variables. Indeed we will consider $\alpha$ as parametrizing a
$\IP^1$ base, over which $2$-folds parametrized by $z_1$ and $z_2$
are fibered.  We then identify $2$-cycles in this $2$-fold.

\noindent Making the change of variables in $\Omega$, one obtains:
\eqn\chvars{\int \Omega ~=~ \int {\psi ~d\alpha~z_1^2 d z_1~dz_2
\over 2 \sqrt{ (1 + \alpha^3 + \coeff{1}{4} (\psi z_2)^2
\alpha^2)~z_1^6 ~+~ \rho^6}}\ ,}
where
\eqn\rhodefn{\rho^6 ~=~ z_2^6 ~-~ 1 \ .}
Considered as a function of $z_1$, $\Omega$ has six
square root branch points.  Integrate $\Omega$ along any contour,
$\gamma_1$, that has $z_1$ large enough so as to encircle all these
branch points.  This contour can then be deformed out to infinity
so as to give
\eqn\intonedone{{1 \over 2 \pi i} ~\oint_{\gamma_1} d z_1~
\Omega ~=~ { \psi ~d\alpha~dz_2 \over 2 \sqrt{ (1 + \alpha^3 +
\coeff{1}{4} (\psi z_2)^2 \alpha^2)}} \ .}
There are two natural integrals over $z_2$, and these
correspond to integrals over two different types of $2$-cycle:
\item{(i)} Integrate $z_2$ between any two zeroes of
\rhodefn.  This means that the $z_1$-contour then sweeps out
a homology $2$-sphere.
\item{(ii)}  Integrate $z_2$ along any contour,
$\gamma_2$, around the
branch points of \intonedone, and thus the $z_1$ and $z_2$
integration sweep out a $2$-torus.

We consider the second possibility first: The integral
over $z_2$ is elementary and yields
\eqn\inttwodone{{1 \over (2 \pi i)^2} ~\oint_{\gamma_2} dz_2~
\oint_{\gamma_1} dz_1~\Omega ~=~ {1 \over 2}~{d \alpha \over
\alpha} \ .}
Note that the factors of $\psi$ have cancelled.  The last
integral is then taken about $\alpha = 0$ and gives
\eqn\firstint{{2 \over  (2\pi i)^3} ~\int ~\Omega ~=~ 1 \ ,}
which reproduces the constant period in \sbp.

To perform the second integral one can perform the integration
directly, but it is first instructive to note that
\eqn\intydone{{d \over d \psi}~ { \psi \over \sqrt{ (1 + \alpha^3 +
\coeff{1}{4} (\psi z_2)^2 \alpha^2)}} ~=~ {d \over d z_2}~
{z_2 \over \sqrt{ (1 + \alpha^3 +  \coeff{1}{4} (\psi z_2)^2
\alpha^2)}} \ .}
Hence if one differentiates the integral of interest with
respect to $\psi$, then the $z_2$-integration becomes trivial:
and the result is the difference, ${\cal I}_k - {\cal I}_j$,
where
\eqn\calIdefn{{\cal I}_k ~=~ {\zeta_k \over \sqrt{ (1 +
\alpha^3 + \coeff{1}{4}(\psi \zeta_k)^2 \alpha^2)}} \ ,}
and $\zeta_k = e^{ \pi i k/3} $.  The remaining integration
over $\alpha$ is a standard elliptic integral of the
holomorphic differential, $d\alpha/\beta$, over the torus
\eqn\auxtorusa{\beta^2 ~=~ \alpha^3 ~+~ \coeff{1}{4}
(\psi \zeta_k)^2 \alpha^2 ~+~ 1 \ ,}
which is equivalent to $P_{E_8}=0$ in \tori.

There are two choices for this integral, the $a$-cycle or the
$b$-cycle.  To reconstruct the original integral of
$\Omega$ over the corresponding $3$-cycle, all we have to do
is integrate these period integrals with respect to the
modulus, $\psi$.  In performing this integral with respect to
$\psi$, the constant of integration is determined by the fact
that for $\psi \to 0$, the integrand of the $z_2$-integral,
$\psi~(1 + \alpha^3 + \coeff{1}{4} (\psi \zeta_k)^2
\alpha^2)^{-1/2}$, vanishes linearly in $\psi$, while the range of
the
$z_2$-integration remains fixed (and finite).  Hence these
integrals vanish linearly in $\psi$ as $\psi \to 0$.  (The first
integral \firstint\ does not vanish in this limit since the
range of integration grows linearly with $\psi$.)
Thus we see that the integral of $\Omega$ described in (i) reduces
to evaluating the periods of a Seiberg-Witten differential
associated with the torus \auxtorusa.

The forgoing procedure is, of course, equivalent to
integrating \intonedone\ with respect to $z_2$
between limits that are distinct values of $\zeta_k$.
Doing this explicitly, one reduces the period integrals
of $\Omega$ to differences of integrals of:
\eqn\lamSWa{\lambda_k ~=~ {1 \over 2}~\log~\left[
{\sqrt{1 + \alpha^3 + \coeff{1}{4} (\psi \zeta_k)^2 \alpha^2} ~+~
\coeff{1}{2} \psi \alpha  \zeta_k  \over \sqrt{1 + \alpha^3 +
\coeff{1}{4} (\psi \zeta_k)^2 \alpha^2} ~-~ \coeff{1}{2} \psi
\alpha  \zeta_k } \right]~{d \alpha \over \alpha} \ ,}
around the cycles of the corresponding tori \auxtorusa. One can check
that indeed $\theta\cdot \lambda_k \sim {d\alpha\over\beta}$ up to
exact pieces, which proves \thetamap. Note also that $\lambda_k$
vanishes for $\psi=0$, which will prove important later.

We now briefly turn to the remaining cases.
The computation for the surface $W_{E_7}=0$, or equivalently
\eqn\singulb{y^2 ~=~ w^4 ~+~ z_1^4 ~+~ z_2^4  ~-~
{1 \over x^4} ~-~ \psi\, w x y z_1 z_2 \ ,}
proceeds in an almost identical manner, except that
slices the surface by hyperplanes, $w = \alpha z_1$,
instead of quadrics.  The integral
over the $2$-torus as in (ii) above gives the result
\firstint. Integrating over $2$-spheres as in (i) above,
leaves the differential:
\eqn\lamSWb{\lambda_k ~=~ {1 \over 2}~\log~\left[
{\sqrt{1 + \alpha^4 + \coeff{1}{4} (\psi \zeta_k)^2~\alpha^2} ~+~
\coeff{1}{2} \psi \alpha  \zeta_k  \over \sqrt{1 + \alpha^4 +
\coeff{1}{4} (\psi \zeta_k)^2~\alpha^2} ~-~ \coeff{1}{2} \psi
\alpha  \zeta_k } \right]~{d \alpha \over \alpha} \ .}
where $\zeta_k = e^{\pi i k/2}$.  This is
to be integrated over the $a$-cycles and $b$-cycles of the
associated torus
\eqn\auxtorusb{\beta^2 ~=~ \alpha^4 ~+~ \coeff{1}{4}
(\psi \zeta_k)^2~\alpha^2 ~+~ 1 \ ,}
which is the same as $P_{E_7}=0$ in \tori.
Finally note that the the analysis for the surface
$W_{E_6}=0$, or equivalently,
\eqn\singulc{ y^3 ~+~ w^3 ~+~ z_1^3 ~+~ z_2^3 ~-~
{1 \over x^3} ~-~ \psi\, w x y z_1 z_2 ~=~ 0 \ ,}
is a little more complicated.  One first makes a shift
$w \to w - y$, followed by a shift $y \to y + {1 \over 2} w$.
The surface then becomes:
\eqn\singuld{ ( 3 w + \psi x z_1 z_2 )~ y^2 ~+~ \coeff{1}{4}
w^3 ~+~ z_1^3 ~-~ \coeff{1}{4}  \psi x w^2 z_1 z_2 ~+~
z_2^3 ~-~ {1 \over x^3}  ~=~ 0 \ .}
In the patch $x = 1$, the $3$-form, $\Omega$, takes the form
\omreduce, with
\eqn\rdefn{ r ~=~ 2 \sqrt{-( 3 w + \psi z_1 z_2)~
(\coeff{1}{4} w^3 ~+~ z_1^3 ~-~ \coeff{1}{4}
\psi w^2 z_1 z_2  ~+~ z_2^3 ~-~ 1)} \ . }
The computation now follows the usual pattern.
Setting $w = \alpha z_1$, and changing variables from
$w$ to $\alpha$, one can then integrate $z_1$ around a contour,
$\gamma_1$, around both cuts in the $y$-plane.  This leaves an
integral of
\eqn\singuld{ {\psi~d \alpha~d z_2 \over \sqrt{(3 \alpha +
\psi z_2)~(\psi z_2 \alpha^2 ~-~ \alpha^3 ~-~4)
}} \ . }
The denominator of the integrand is once again the square root
of a quadratic in $z_2$.  The Seiberg-Witten differential is:
\eqn\lamSWc{\lambda_k ~=~ {1 \over 2}\,\log\,\left[ {\psi \zeta_k
\alpha^2 + (\alpha^3 - 2) ~+~ \alpha~\sqrt{( 3 \alpha + \psi
\zeta_k)~(\psi \zeta_k \alpha^2 - \alpha^3 - 4)} \over \psi \zeta_k
\alpha^2 + (\alpha^3 - 2) ~-~ \alpha~ \sqrt{( 3 \alpha + \psi
\zeta_k )~(\psi \zeta_k \alpha^2 - \alpha^3 - 4)} } \right]~
{d \alpha \over \alpha}}
on the Riemann surface
\eqn\auxtorusc{\beta^2 ~=~ ( 3 \alpha + \psi \zeta_k)~(\psi
\zeta_k \alpha^2 - \alpha^3 - 4) \ .}
Upon simply substituting $\beta = 2 (3 (w+y)  +
\psi z_1 \zeta_k) (y - w)/z_1^2$ and $\alpha = (w + y)/z_1$,
this curve is easily seen to be equivalent to the torus
$P_{E_6}=0$ in \tori.

We did not try to perform an analogous computation for the $E_5$
potential in \delPezz, but have no doubt that it would work out in a
similar way.

One should note that there are some fundamental differences
between the foregoing calculations and the usual calculation of
Seiberg and Witten.  Apart from the vanishing of
\lamSWa\ at $\psi = 0$, which is a regular point of the
moduli space, it should be noted that the differential \lamSWa\ is
not single valued.  Logarithms certainly arise in some of the
standard forms of $\lambda$, but usually these logarithms
can be removed by suitable integrations by parts.  Put another
way, the branches of the logarithm play no role because the shift
of $2 \pi i$ associated with crossing the cuts merely
adds the (cohomologically trivial) derivative of a
meromorphic function to $\lambda$.  In the foregoing,
this ambiguity in the logarithm adds multiples of $2 \pi i
(d \alpha/\alpha)$ to $\lambda_k$, and this has residues
at $\alpha = 0, \infty$.

This result is somewhat new from the one-dimensional
perspective of the Seiberg-Witten differential, but it is
easily understood form the perspective of the $3$-fold.
There is an obvious ambiguity in the integrations of
type (i) between the zeroes of \rhodefn:  the path can
wind around the set of all branch points of the integrand
of \intonedone, exactly as in (ii).  This adds constant
($2 \pi i$ times an integer) shifts to the periods of
$\lambda_k$, {\it i.e.} to both $\phit$ and $\phitd$.

However, from the one-dimensional perspective, while there
is only one torus, there are infinitely many Seiberg-Witten
differentials, $\lambda_{(n)}$, which differ only by
multiples of $d \alpha/\alpha$ so that $\lambda_{(n)}$ has
residue $2 \pi i n$ at $\alpha = 0$.  If one now tries to use
the Seiberg-Witten differential as a metric, as in \KLMVW,
then one must take infinitely many copies, indexed by $n$,
of the basic torus, with a metric obtained from $\lambda_{(n)}$
on the $n^{\rm th}$ copy.  This foliation of the basic torus
is sewn together at the logarithmic branch cuts of
$\lambda_{(n)}$.  For the differential \lamSWa, these cuts
are located at $\alpha= e^{2 \pi i/3}$.  BPS geodesics can thus
find their way between sheets by cycling around these branch
points, and thus the counting of BPS states \KLMVW\ will be very
different as compared to ordinary \nex2 Yang-Mills theories.

\newsec{Periods, prepotentials and monodromies from the $E_8$ torus}

In order to obtain explicit expressions for the geometric
del Pezzo periods, $\tilde\Pi_i$, we can now simply make use
of the hypergeometric systems \pfI\ related to the SW tori \tori.
The correct integral symplectic
linear combinations of the solutions can be simply determined
by matching the asymptotic expansions with the period integrals
over the differentials $\lambda$ of the previous section.

To be specific, we will consider here only the
model $\cB_8$; the other theories can be treated in an analogous way.
For $\lambda$ is as in \lamSWa, we find,
in terms of the inverse variable $\u\equiv -1/(432\ca)$
and a standard homology basis $\gamma_{a,b}$ on the $E_8$
torus, the following expressions for the periods:
\eqn\integper{
\eqalign{
\phit(\u)\ &=\   {3\over \pi^2}\int_{\gamma_a}\lambda\ =\
\Big(\int\varpi(\u){d\u\over \u}\Big) + \delta \cr
\phitd(\u)\ &=\ {3\over \pi^2}\int_{\gamma_b}\lambda\ =\
\Big(\int\varpi_D(\u){d\u\over \u}\Big)+ \delta_D \ .
}}
Here, $\delta,\delta_D$ are integration constants, and
\eqn\torusper{
\eqalign{
\varpi_D(\u)\ &=\ {3^{1/4}\over 4\pi^{3/2}i}
\Big(\xi F_0(\u) + {1\over \xi} F_1(\u)\Big)\cr
\varpi(\u)\ &=\ {3^{1/4}\over 4\pi^{3/2}i}\
\Big(\rho\xi F_0(\u) +  {1\over \rho\xi} F_1(\u)\Big)
\ ,\cr
}}
with $\xi\equiv
-{{i\,{3^{{1/4}}}}\over
   {{2^{{2/3}}}\,{{\pi }^{{3/2}}}}}{{{\Gamma}({1/3})}^3}$,
$\rho\equiv e^{2\pi i/3}$ and
\eqn\hyper{
F_0(\u)\ =\ {\u}^{1/6}{}_2F_1\big(\Coeff16,\Coeff16,\Coeff13;
\u\big)\ ,
\qquad\
F_1(\u)\ =\ {\u}^{5/6}{}_2F_1\big(\Coeff56,\Coeff56,\Coeff53;
\u\big)\ .
}
The hypergeometric system is known to have singularities at
$\u=0,1,\infty$. With the help of well-known formulas for the
analytic continuation of ${}_2F_1$, we can easily find expansions of
$\phit(\u),\phitd(\u)$ in local patches near these singularities.
However,
when comparing different patches, we must be careful with the
integration constants $\delta,\delta_D$ in \integper. For expansions
around $\u=0$ these constants must be zero, since we know from the
vanishing of $\lambda$ \lamSWa\ for $\u=0$ that both periods $\phit$
and
$\phitd$ vanish there. From this, or from the explicit expression
for $\lambda$, we can match the integration constants in other
patches.

\noindent Explicitly, we find the following local expansions near the
singularities:

\noindent 1) Near $\u=0$, which is the point of intersection with the
theory discussed in \doubref\ganorII\gms:
\eqn\nullexp{
\eqalign{
(2\pi i)\,\phitd(\u) =
-{3^{5/4}\over\sqrt\pi} \u^{1/6}\Big[\,
\xi\big(1+\Coeff1{84}\u+\cO(\u^2)\big)+
{1\over\xi}\u^{2/3}\big(\Coeff15+\Coeff5{132}\u+\cO(\u^2)\big)\,\Big]
}}
and $\phit(\u)=\phitd(u)|_{\xi\to\rho\xi}$. The vanishing of both
$\phit$ and $\phitd$, as well as the absence of a logarithm, reflects
conformal invariance of the theory at $\u=0$. This is similar to the
$SU(2)$ theories with $N_f=1,2,3$ \doubref\witx\TMHS. The
prepotential $\cF_0(\phit)={1\over2}\tau_{FP}\phit^2 -{i\over 10
\sqrt3\xi^6} \phit^6+ \cO(\phit^{10})$ is thus polynomial in $\phit$,
and for the $U(1)$ coupling constant \taudef\ at the origin we have:
$\tau_{FP}=\rho^2$. This is, as expected, a fixed point of the
modular
group (this value is consistent with
$\u(\tau)={1\over864}(J(\tau)+\sqrt{J(\tau)(J(\tau)-1728)})$, from
which we find for the other fixed point of the modular group:
$\u(\tau=i)=2$).

\noindent 2) Near $\u=1$:
\eqn\oneexp{
\eqalign{
(2\pi i)\,\phit(\u) &=
{\rm const.}-
\Coeff1{2\pi}\Big[\big\{(\u-1)(1+\log[432])+\cO((\u-1)^2)
\big\}-
\cr&\qquad\qquad
(i-i\,\phitd(u))\log[\u-1]\Big]
\cr
(2\pi i)\,\phitd(\u) &=  2 \pi i + i(\u-1) -\Coeff{31}{72}i(\u-1)^2
+\cO((\u-1)^3)
}}
This means that at $\u=1$ there is a conifold singularity
that is analogous to the SW monopole singularity.
It corresponds to a massless BPS state
with quantum numbers $(n_0,n_2,m_2)=(-1,0,1)$.

\goodbreak
\noindent 3) Near $\u=\infty$:

For these expansions we go back to the original variable,
$\ca=-1/(432\u)$. From the analytic continuation we have the
following integration constants:
$$
\delta=  \Coeff12-{1\over2\pi i} \log[432]\,,\ \ \ \
\delta_D= \Coeff{13}{24} +{1\over8 \pi^2 }\log[432]^2\ ,
$$
and thus:
\eqn\infexp{
\eqalign{
(2\pi i)\,\phit(\ca) &=  \log[\ca]-60\ca+6930\ca^2+\cO(\ca^3)\cr
(2\pi i)\,\phitd(\ca) &=
\Coeff56i\pi-\big((30+126\Coeff i\pi)\ca+\cO(\ca^2)\big)
+\cr&\qquad
\big(\Coeff12-30\Coeff i\pi \ca +\cO(\ca^2) \big)\log[\ca]
-\Coeff1{4\pi i} \log[\ca]^2\ .
\cr}}
The SW period $\phitd(\ca)$ indeed reproduces the three-fold period
$-\Pi_3$, given in \ppII\ and (A.2) below, when expressed in terms
of $t_S\equiv \phit(\u)$. Consequently, up to a constant, the
prepotential near $\u=\infty$ is:
\eqn\Finf{
{\cF}_\infty(\phit)={1\over6}\phit^3+
{1\over4}\phit^2-{5\over12}\phit+
{1\over8 \pi^3 i}\sum_{k=1}^\infty\, n_k^r\, Li_3(e^{2\pi i k\,
\phit})\ ,
}
where the instanton coefficients $n_k^r=\{252,-9252,..\}$
are as in \yuk, and $Li_3$ is the
trilogarithm \HM. This rigid prepotential is obviously quite
different
from the usual SW theory prepotentials,
${\cal F}_\infty(a)\sim a^2 \log a^2 + \cO(1/a)$ for large $a$,
and also from that of a higher-dimensional field theory including
the effects of Kaluza--Klein excitations. We stress again that
\Finf\ reflects the behaviour of the rigid sub-moduli problem
at the large complex structure point of the threefold.
This is different than in usual field theory
embeddings in string theory, where the ``point at infinity'' of
the field theory does not coincide with the ``point at infinity''
of the string theory.

\noindent  From the above local expansions we find the following
monodromy matrices:
\eqn\monmat{
M_0= \pmatrix{ 1 & 0 & 0 \cr 0 & 0 & -1 \cr 0 & 1 & 1 \cr  }
\ ,\ \
M_1=\pmatrix{ 1 & 0 & 0 \cr -1 & 1 & 1 \cr 0 & 0 & 1 \cr  }
\ ,\ \
M_\infty=\pmatrix{ 1 & 0 & 0 \cr -1 & 1 & 0 \cr 0 & 1 & 1 \cr  }\ ,
}
which act on $\tilde\Pi=(1,\phit,\phitd)$ and which represent the
quantum symmetries of the theory. They obey $M_1\cdot M_0=M_\infty$,
${M_0}^6=\bfone$, and we can easily identify the modular group
$SL(2,\ZZ)$ in the lower right block, i.e.., $M_0\sim T^{-1}S$,
$M_1\sim TS^{-1}T$ and $M_\infty\sim T$. From these generators we can
build pure shifts: $\phit\to \phit\pm1$ or $\phitd\to \phitd\pm1$.
These can be subtracted from the above generators, and thus the full
duality group is given by a semi-direct product of $SL(2,\ZZ)$ with
$\ZZ\times \ZZ$.

\newsec{Discussion}

In this paper we have investigated del Pezzo singularities of
Calabi-Yau three-folds in \nex2 supersymmetric
type IIA string compactifications. We have
isolated a one-dimensional moduli space, $\cM(\cB_n)$, which
intersects the large complex structure point of the three-folds. This
moduli space corresponds to a closed sub-monodromy problem, and
should therefore be associated with a consistent sub-theory of the
full string compactification.

The question arises as to what the structure of this consistent
sub-theory is, and what precise physical meaning the moduli
space has. This is like
asking what the physical meaning of the moduli space of a
Seiberg-Witten curve is, without a priori knowing the underlying
Yang-Mills theory. In other words, we need to ask what the physical
system is whose moduli space is the moduli space of a del Pezzo
surface.

Clearly, although the theory has six non-compact dimensions
(since we used a non-compact threefold to model it),
it does not describe a six-dimensional Lorentz invariant theory.
Moreover the effective description in terms of special geometry
is an essentially four dimensional concept. However, the situation
should be compared to the effective theory obtained in the limit of
a large base $\IP^1$ in a K3 fibered Calabi--Yau manifold:
although the global geometry
does not allow to recover precisely the six-dimensional space-time
theory of the pure $K3$ compactification, one does recover the
precise moduli space of the $K3$ compactification, that is, the
singularities, monodromies and gauge coupling dependence. This
is because the $\IP^1$ becomes locally flat and
the local properties of the theory become indistinguishable
from that of the truly six-dimensional theory.

{}From the discussion in sections 2 and 3 we know that the coordinate
$\ca$ of the moduli space describes essentially the string tension
of the six-dimensional non-critical $E_n$ string. In particular
there are no relevant geometrical parameters to vary. It is
therefore suggestive to interpret this moduli space primarily as
the moduli space of the six-dimensional string theory itself,
and not of a  particular compactification.
Specifically, its singularities
should be related to the properties of the string
theory itself, and not just to an interplay with the geometry of the
compactification. The behaviour at $\ca=\infty$, where the
string tension vanishes, is clearly consistent with this picture.
The large complex structure point, $\ca=0$, is stringy either:
the behaviour of the theory at this point is
characterized in terms of the
type II A string world-sheet instantons expansion. These instantons
are supported by the same rational curves which give also
rise to the BPS {\it states} of the NCS relevant at the $\ca=\infty$
singularity. On the other hand, the third singular point,
the conifold singularity, is slightly
different in that it needs a non-vanishing value of the $b$ field
in the directions in which Lorentz invariance is broken, and would
therefore be absent in a Lorentz invariant theory.

In order to gain a better understanding of
the structure of the theory,
it would be interesting to repeat
the conformal field theory analysis of \ovI\ in the present context.
It would also be important to understand the
precise counting of the BPS states, as well as their underlying
algebraic structure. For the counting, the SW formulation involving
the peculiar differentials $\lambda$  might be very useful. As for
the
algebraic structure, recall that the homology of the del Pezzo
surfaces is dim$H_*(\IB_n,\ZZ)=(b_0=1,b_2=n+1,b_4=1)$, which leads to
a Lorentzian period lattice of type $\Gamma_{n+2,2}$. $H_2$ forms a
sub-lattice of type $\Gamma_{n+1,1}$, which is related to the period
lattice \Arn\ of the relevant simple elliptic singularity in \tori\
(note that we considered in this paper only a one-dimensional
sub-lattice of $H_2$). We expect the BPS states to form a generalized
Kac-Moody algebra \HM\ associated with the period lattice, and the
prepotential $\cal F$ and the topological free energy ${\cal F}_1$ to
be
associated with the corresponding automorphic forms.

Finally, it would also be instructive to investigate the precise
relationship of the \nex2 supersymmetric theories discussed in this
paper with the $E_n$ superconformal fixed points of \JMDN. These are
based on the simple singularities of type $E_n$, while the theories
discussed in the present paper are related to the elliptic
singularities of type $\hat E_n$. It is well-known from the adjacency
diagrams of singularity theory \SaitoES\ that the $E_n$ singularities
can be obtained by deformations of the $\hat E_n$ singularities. Both
types of theories have superconformal points in their moduli spaces,
and we suspect these to belong to different universality classes.
Very crudely speaking, the $\hat E_n$ superconformal fixed point
might be a stringy, ``affine'' version of the $E_n$ superconformal
fixed point.

We believe it would be important to shed more light on these issues,
and intend to report on them in the future.
As a remark, note also in this context that we can treat
the conifold singularity \cof\ in a very similar way as we did for
the del Pezzo singularities. Since this singularity is much more
generic than the del Pezzo singularities, we might gain considerable
understanding of non-perturbative string theory by studying this
theory in more detail.

\goodbreak
\vskip2.cm\centerline{{\bf Acknowledgements}}
\noindent
We would like to thank J.\ Minahan,
D.\ Nemeschansky, S.\ Theisen and C.\ Vafa for valuable discussions.
N.W. is supported in part by funds provided by the DOE
under grant number DE-FG03-84ER-40168.

\vfill\break
\appendix{A}{Prepotential and periods of the three-fold $\xf$}

The elliptically fibered three-fold $\xf$ with base $\IF_1$ has hodge
numbers $h^{1,1}=3,\ h^{1,2}=243$. Its description in
terms of toric geometry and its phases has been discussed in much
detail
in \refs{\mv{,}\candI{,}\ste{,}\KMV}.

The four-dimensional $N=2$ vector moduli space is described by the
holomorphic prepotential of special geometry $\cx F(t_i)$. For $\xf$
it takes the following asymptotic form in the two phases, Phase $I$
and
Phase $II$:
\eqn\ppI{\eqalign{
\cx F_I(t_i) =&
-\h  \tef^2 \tee-\tef \teb \tee-{3\over2} \tef \tee^2-\teb \tee^2
-{4\over3} \tee^3
\cr&+\h
 \tef \tee+{3\over2} \tef+\teb+{23\over6} \tee -c_0\ , \cr
\cx F_{II}(t_i) =&
-\h  \tfp^2 \tbp-\h  \tfp^2 \tep-{3\over2}
 \tfp \tbp^2-3 \tfp \tbp \tep-{3\over2} \tfp
 \tep^2-{3\over2} \tbp^3\cr&
-{9\over2} \tbp^2 \tep
-{9\over2} \tbp \tep^2-{4\over3} \tep^3+\h  \tfp
\tbp+\h  \tfp \tep+{1\over4} \tbp^2\cr&
+\h  \tbp \tep
+{3\over2} \tfp+{17\over4} \tbp+{23\over6}
\tep-c_0 \ ,
}}
where $c_0=-i\zeta(3)/2(2\pi)^3\chi$. The primed coordinates in
Phase $II$ are related to those in Phase $I$ by $\tbp=-\teb,\
\tfp=\tef+\teb, \tep=t_S=\tee+\teb$. The volume of the 4-cycle is
$t_S$, while the volume of the 2-cycle shrinking at the
flop is $\teb$.

The period vector $\Pi$ has entries $(\cx F_0,\ \cx F_i,\ 1,\ t_i)$
in a basis with canonical symplectic metric. Applying an integral
symplectic transformation puts it into the following form:
\def\frac#1#2{{#1 \over   #2}}
\eqn\pvI{
\Pi_{II}=\pmatrix{
\h  \tfp ^2 \tee +{3\over 2}  \tfp  \tee ^2-{1\over 6} t_S ^3+
{3\over 2}  \tee ^3+{3\over 2}  \tfp -{5\over 12} t_S +
{17\over 4} \tee -2c_0\cr
-\tfp  \tee -{3\over 2}  \tee ^2+\h  \tee +{3\over 2}    \cr
\h  t_S ^2-\h  t_S -{5\over 12   }\cr
-\h  \tfp ^2-3 \tfp  \tee -{9\over 2} \tee ^2+\h  \tfp +\h  \tee
+{17\over 4}\cr
1\cr \tfp  \cr t_S  \cr \tee}}
The periods $(1,\ t_S,\ \Pi_3)$ depend then (classically)
only on the K\"ahler coordinate
$t_S$ and define our sub-monodromy problem. Moreover,
$\tee$ is the additional period treated as a background field in
the $\Bp_n$ theory.
\goodbreak
\subsubsec{Elliptic curves}\br

\def\ss#1{{\scriptstyle{#1}}}
{\vbox{\ninepoint{
$$
\vbox{\offinterlineskip\tabskip=0pt
\halign{\strut\vrule#
&\hfil~$\ss{#}$
&\vrule#&~
\hfil ~$\ss{#}$~
&\hfil ~$\ss{#}$~
&\hfil $\ss{#}$~
&\hfil $\ss{#}$~
&\hfil $\ss{#}$~
&\hfil $\ss{#}$~
&\hfil $\ss{#}$~&\hfil $\ss{#}$~&\hfil $\ss{#}$~
&\vrule#\cr
\noalign{\hrule}
&    &&  d & 1 &  2  &  3  &  4 & 5 & 6 & 7 & 8 &\cr \noalign{\hrule}
&n   &&    &   &     &      &       &        &                 &
&&\cr
&8   &&    &
-2&   762&   -246788&   76413073&   -23436186174&
7209650619780&   -2232321201926990&   696061505044554010&\cr
&7   &&    &
0&   3&   -224&   12042&   -574896&   26127574&   -1163157616&
   51336812456&\cr
&6   &&    &
0&   0&   -4&   135&   -3132&   62976&   -1187892&   21731112&\cr
&5   &&    &
0&   0&   0&   5&   -96&   1280&   -14816&   160784&\cr
&0   &&    &
0&   0&   -10&   231&   -4452&   80958&   -1438086&   25301064&\cr
\noalign{\hrule}}
\hrule}$$
\vskip-7pt
\noindent
{\bf Table 1}: Number of genus one curves of degree $d$
for $\cx B_n,\ n=8,7,6,5,0$.
\vskip7pt}}
\noindent
For comparison we have also given the number of
genus one curves for $\cx B_0$ in table 1, which have been
determined in \cand.

\listrefs
\end